Determination of Effective Temperatures and Luminosities for Rotating Stars

A. Gillich, R. G. Deupree, C. C. Lovekin, C. I. Short, and N. Toqué

Institute for Computational Astrophysics and Department of Astronomy and Physics

Saint Mary's University, Halifax, NS B3H 3C3, Canada

bdeupree@ap.smu.ca

ABSTRACT

Spectral energy distributions for models of arbitrarily rotating stars are computed using two dimensional rotating stellar models, NLTE plane parallel model atmospheres, and a code to integrate the appropriately weighted intensities over the visible surface of the stellar disk. The spectral energy distributions depend on the inclination angle between the observer and the rotation axis of the model. We use these curves to deduce what one would infer the model's luminosity and effective temperature to be assuming the object was nonrotating.

*Subject headings:* stars: atmospheres, stars: rotation

1. INTRODUCTION

For spherical stars the relationship between the observed magnitudes and colors and physically meaningful information about the star, such as the luminosity and effective temperature, is relatively straightforward given a reasonable estimate of the distance to the star.

This relationship becomes more complicated for stars that rotate rapidly because the colors and magnitudes may depend appreciably on the inclination between the observer and the star's rotation axis. Thus, even something as prosaic as the location of a star in the HR diagram involves an additional level of complexity because of the dependence on the inclination. Instead of a point in the theoretician's HR diagram, a star is represented by a curve with the inclination determining the specific location on the curve. We must know this curve, which we shall designate the inclination curve, in order to translate what we observe into information that could provide either constraints on or validation of rotating stellar models.

There has been extensive work on the inclination curve (e.g., Collins 1966; Collins & Harrington 1966; Hardorp & Strittmatter 1968; Maeder & Peytremann 1970). These authors used spherical models which include a radial centrifugal force for the stellar structure. The surface distortion is based on a Roche potential and the surface effective temperature as a function of latitude computed with von Zeipel's (1924) law relating the effective temperature to the effective gravity. Generally speaking, for massive stars as the rotation rate increases the polar effective temperature becomes marginally hotter and the observed energy output higher, and the equatorial effective temperature significantly cooler (so that the average effective temperature becomes slightly cooler) and less luminous. The inclination curve becomes progressively longer as the rotation rate increases and hence the need to determine where on the inclination curve an observed star may be located becomes more critical.

This work has been recently built upon by Lovekin, Deupree, and Short (2006, hereinafter LDS) who used fully two dimensional models of rotating stars and the PHOENIX

NLTE model atmosphere code (Hauschildt & Baron 1999) to compute the spectral energy distributions of rotating stars. The spectral energy distribution is the weighted sum of the radiative intensities from the surface in the direction of the observer, integrated over the visible surface of the star. The method for performing these calculations is effectively the same as that used by current researchers (e.g., Slettebak, et al. 1980, Linnell & Hubeny 1994, Fremat, et al. 2005), although different from the previously used limb and gravity darkening calculations (e.g., Reiners & Schmitt 2002, Reiners 2003, Townsend, Owocki & Howarth 2004). The objective of LDS was to determine if the spectral energy distribution contained clues which depend on the angular momentum of the model. LDS focused on a limited wavelength range of the spectral energy distribution and used the shape to determine the deduced effective temperature as a function of the inclination. Here we wish to extend this work by computing the spectral energy distribution over much more of the wavelength range. The effective temperature will still be determined by the shape of the spectral energy distribution, while the luminosity will be calculated from the area under the spectral energy distribution curve. This would give us the complete inclination curve.

In § 2 we shall describe the calculations required to produce the spectral energy distribution of a rotating star. The spectral energy distributions are presented in § 3 and the deduced inclination curves corresponding to various rotation velocities for both uniformly and differentially rotating models are discussed in §4 and §5, respectively.

## 2. METHOD OF CALCULATION

There are three steps required for computing the spectral energy distribution of rotating stars. First we must compute the stellar structure of the rotating model, requiring the solution of equations for hydrostatic equilibrium, energy conservation, along with Poisson's equation for the gravitational potential for an imposed specification of the rotation rate as a function of location inside the model (the rotation law). The only information from these models used for the computation of the inclination curve is the effective temperature, effective gravity and surface radius as functions of latitude. The effective temperature and radius are computed directly by the model and the effective gravity is computed afterwards from the spatial distribution of the gravitational potential and the rotation law.

To perform the integral to compute the spectral energy distribution as a function of inclination, we need much higher resolution than the 2D code needs to solve the structure. This higher angular resolution is required for getting the direction to the observer and hence the intensity in that direction at each point on the surface with sufficient accuracy. The quantities we require can easily be obtained on this finer grid. In practice, we have found that a grid of 200 latitudinal and 400 azimuthal points gives more than adequate resolution. This was tested by the similarity in the spectral energy distribution at all inclinations for spherically symmetric models. We assume that the emergent intensity from each point on the grid can be represented by the emergent intensity of a plane parallel model atmosphere with the given effective temperature and effective gravity. The emergent intensity will be a function of the angle from the normal to the surface at that point. The plane parallel assumption should be an adequate approximation as long as the horizontal structure does not change much over a photon mean free path. This should be true for uniformly rotation models with the possible exception of near the equator for models rotating near critical rotation, when the effective gravity is low and the curvature in units of the photon mean free path may be significant. The equatorial region is only a modest contributor to the flux, and we shall assume the plane parallel approximation holds for the equatorial region as well as elsewhere. This argument is essentially that made by Hardorp & Strittmatter (1968). The plane parallel approximation may also fail for strongly differentially rotating models which are rotating rapidly because these models may develop bulges on the surface which could be seen from distant surface locations. These bulges would appear symmetric about the local surface normal only for points on the polar axis because of the azimuthal symmetry. We shall not include any models with sufficient bulges for which this could be the case.

We have computed a grid of plane parallel model atmospheres for a range of effective temperatures and gravity. At each point on our finely zoned stellar surface, we compute the angle between the local normal to the surface and the inclination and interpolate within the model atmosphere grid as functions of effective temperature, effective gravity and this angle. The result is the emergent intensity coming from this point on the surface in the direction of the observer. The third step of this process performs this interpolation and computes the weighted integral of this intensity over the visible stellar surface to produce the flux the observer would see. This is performed over many wavelengths to generate the observed spectral energy distribution. All our work here computes the observed spectral energy distribution at inclinations between 0° and 90° in 10° increments.

The surface location and the values of the effective temperature and effective gravity everywhere on the surface are computed with the 2.5 dimensional stellar evolution code ROTORC (Deupree 1990, 1995, 1998). The half-dimension means that there is a conservation equation for the azimuthal component of the momentum, but that the model has azimuthal symmetry. Here we shall be concerned only with ZAMS models of B stars, for which we impose an angular momentum distribution and then solve for the density, temperature, and gravitational potential distributions. The composition we take to be uniform with X=0.7, Z=0.02. The independent coordinates are the fractional equatorial surface radius and the colatitude. Equatorial symmetry is assumed here, but only for computational resolution reasons. The calculations are performed in the inertial frame. The shape of the surface, as given by the fractional equatorial radius as a function of latitude, is found by requiring that it be an equipotential surface, while its location, which requires the surface equatorial radius, is determined by the requirement that the model contain the total mass imposed. The equipotential assumption is imposed only because we need some method for determining what the shape of the surface is; we do not require that the variables be constant on this surface.

We have two boundary conditions we need to apply on the surface boundary. The simplest conditions, the so called "zero boundary conditions" (Schwarzschild 1958) are often adequate for massive main sequence and have been used in other models of rotating stars (e.g., Sackmann & Anand 1970; Jackson, MacGregor & Skumanich 2005). A more realistic alternative is to match all the dependent variables at some location with those generated from a code that specifically models the surface layers (e.g., Paczynski 1970, Kippenhahn & Weigert 1990). This may be merely a stellar atmospheres code, a code that integrates the stellar structure equations inward from the surface to a specified interior mass, or a combination of both. This approach poses a problem in rotating stars because von Zeipel's law indicates that the surface temperature should be independent of latitude because it is on an equipotential surface, but that the effective temperature can vary on the order of a factor of two between the pole and equator for sufficiently rapidly rotating stars. Stellar atmospheres models will produce a dependence between the surface temperature and the effective temperature which von Zeipel's law does not allow. This raises the question as to whether stellar atmosphere models based effective temperatures and effective gravities obtained from von Zeipel's law can be used at all to compute line profiles and spectral energy distributions of rapidly rotating stars. We shall discuss some of the possibilities for rotating model structures once we outline our surface boundary condition.

The surface boundary condition we impose on the temperature is more realistic than the zero boundary conditions but simpler than the matching of a stellar atmosphere model with the interior structure model. Part of the reason is that ultimately we wish to be able to follow the flow patterns generated in rotating stars and need to be able to compute the appropriate conservation laws all the way to the surface. The energy conservation equation we are solving in the outer layers of these static, ZAMS B stars is that the divergence of the radiative flux vanishes. We use the diffusion approximation relating the radiative flux to the temperature gradient so that we have a second order differential equation in two spatial dimensions. Azimuthal symmetry and equatorial symmetry provide the horizontal boundary conditions, and a third is specified by the vanishing of the radial gradient of the temperature at the center of the model. The fact that the energy conservation equation is more complicated there because of convection and nuclear energy generation does not alter this detail. The remaining boundary condition is specified at the model surface. The surface flux is taken to be $2\sigma T_{surf}^4$, where $T_{surf}$ is the temperature in the last radial zone at a given angle, and we assume that the optically thin temperature distribution is that of a simple gray atmosphere:

$$T^4(\tau) = \tfrac{3}{4} T_{eff}^4 \left(\tau + \tfrac{2}{3}\right) \qquad (1)$$

Applying this relation between the surface flux and the surface temperature at each latitude in the finite difference expression of the energy conservation equation provides the remaining condition that allows us to determine the temperature everywhere. Of course the energy equation is only one of the conservation and other (e.g., Poisson's equation for the gravitational potential) coupled equations which are solved simultaneously with the Henyey method. Once we have the surface temperature, we obtain the effective temperature at each latitude by multiplying the local surface temperature by the fourth root of two.

These considerations remind us that we do not have a completely self consistent model for the structure of rotating stars. In von Zeipel's paper, a structure could be found only for an artificial energy generation rate, from which Eddington (1925) concluded not only that a star rotating uniformly, which can be generalized to a conservative rotation law, cannot be simultaneously in hydrostatic and radiative equilibrium but also that meridional circulation must result. This led to attempts to determine the meridional flow using perturbation analysis (e.g., Sweet 1950, Mestel 1966). An unfortunate result of this approach is that the flow velocities near the surface are proportional to the inverse of the density and hence very large. Tassoul and Tassoul (1995) have shown that this problem disappears if an eddy viscosity associated with small scale turbulence acting on the large scale flow is included. Steady state solutions which include meridional circulation and small scale turbulence by Lara and Rieutord (2007) produce a non conservative rotation law (so the hypotheses of von Zeipel's law are not met), a baroclinic structure near the surface, and a significantly reduced ratio of the polar to the equatorial radiative flux near the surface in comparison to that of von Zeipel's law. We note that near the surface both a baroclinic structure and reduced flux ratio hold in our calculation as well, although we make no claim at present that this is more than a coincidence.

It is possible that the final state could be one in which the meridional circulation redistributes the angular momentum in such a way that the meridional circulation disappears. Uryu and Eriguchi (1995) investigated this by imposing azimuthal symmetry, the rotational velocity on the equator at all depths, and the appropriate boundary conditions and solving radiative equilibrium and two equations based on the radial and latitudinal components of hydrostatic equilibrium to determine the rotation law as well as the density and temperature throughout the model. There is no reason for this to be a conservative law, and it is not.

Arguments have been made (e.g., Osaki 1966, Smith 1970, Smith & Worley 1974) that the final result for a star with a conservative rotation law in the interior could be an atmosphere in which the rotation and circulation adjust in such a way so as to maintain von Zeipel's law. It is perhaps noteworthy that this conclusion was made at a time when it was thought that the surface circulation velocities would be large and that this is not the result of the calculations of Lara and Rieutord (2007).

Despite past and current efforts the exact structure and rotation law near the surface of rotating stars with radiative outer layers is not known. It is our intention with the 2D evolution code to be able to simulate the time dependent behavior of the flow generated by these initial models to determine how the rotation law and structure evolve.

In this paper we shall be dealing with conservative rotation laws, but both uniform and differential rotation are included. The total potential, $\Psi$, may be written as the sum of the gravitational potential, $\Phi$, and the rotational potential (e.g., Tassoul 2000)

$$\Psi = \Phi - \int_0^\varpi \Omega^2(\varpi')\varpi' d\varpi' = \Phi - \frac{\Omega^2 \varpi^2}{2} + \int_0^\varpi \varpi'^2 \Omega(\varpi')\frac{d\Omega(\varpi')}{d\varpi'} d\varpi' \quad (2)$$

where $\varpi$ is the distance from the rotation axis and $\Omega$ is the rotation rate. The second equation on the right possesses some numerical advantages (Toqué, et al. 2007). The last integral on the right was missing from LDS, causing too much flux near the pole in their differentially rotating models. The effective gravity can be calculated from the derivative of the total potential taken normal to the surface. The effective temperature, effective gravity, and radius, all as functions of latitude, make up the required input for the integration of the model atmosphere properties over the visible disk of the model.

The model atmospheres are computed with the PHOENIX code (Hauschildt and Baron 1999). In addition to the composition, the input requirements for these plane parallel atmospheres are the effective temperature and the gravity. One of the more attractive features of PHOENIX is the capability to include a large number of energy levels for several ionization stages of many elements in NLTE (Short, et al. 1999). Table 1 presents the number of energy levels and line transitions for the ionization stages of the elements included. For those species treated in NLTE, only energy levels connected by transitions for which log (gf) is greater than -3 in the PHOENIX line list are included in the statistical equilibrium equations. All other transitions for that species are calculated with occupation numbers set equal to the Boltzmann distribution value with the excitation temperature equal to the local kinetic temperature, multiplied by the NLTE departure coefficient for the ground state in the next higher ionization stage. We discovered that the inclusion of silicon and phosphorus in NLTE led to convergence difficulties for models with effective temperatures less than 26000K, so we forced these elements to be in LTE for models with effective temperatures of 24000K and below.

The energy level and bound - bound transition atomic data have been taken from Kurucz (1994) and Kurucz & Bell (1995). The resonance-averaged Opacity Project (Seaton, et al. 1994) data of Bautista et al. (1998) have been used for the ground state photoionization cross sections of Li I-II, C I-IV, N I-VI, O I-VI, Ne I, Na I-VI, Al I-VI, Si I-VI, S I-VI, Ca I-VII, and Fe I-VI. For the ground states of all stages of P and Ti and for the excited states of all species, we used the cross-sectional data previously incorporated into PHOENIX from either Reilman & Manson (1979) or from the compilation of Mathisen (1984). The coupling among all bound levels by electronic collisions is calculated using cross sections calculated from the formulae given by Allen (1973). The cross sections of ionizing collisions with electrons are calculated from the formula of Drawin (1961). Further details are provided by Short et al. (1999).

One of the reasons for including an extensive suite of NLTE energy level populations is the desire to calculate line profiles as well as SEDs with as much realism as possible. The analysis of line profiles is outside the scope of the current work, but will be considered in the near future.

With the rotating model and model atmospheres calculated, we are ready to integrate the intensity in the direction of the observer over the observable disk to obtain the flux at a given distance, d, from the star:

$$F_\lambda = \int_\theta \int_\varphi \frac{I_\lambda(\xi(\theta,\varphi))}{d^2} dA_{proj} \qquad (3)$$

where $F_\lambda$ is the flux at wavelength $\lambda$, $\xi$ is the angle between the local surface normal at location $(\theta,\varphi)$ on the surface and the observer, $I_\lambda$ is the intensity , and $dA_{proj}$ is the projected surface area element as seen by the observer. The detailed geometry of the integration is presented by LDS. We obtain the deduced effective temperature, $T_{ed}$, from the shape of the spectral energy distribution, as we outline in the next section, and the deduced luminosity, $L_d$, from the total flux:

$$L_d = 4\pi d^2 \int_0^\infty F_\lambda d\lambda \qquad (4)$$

Both of these quantities are functions of the inclination. We note that this approach automatically includes the effects of limb darkening and gravity darkening built in to the calculation without the need of somewhat arbitrary prescriptions for them which can significantly affect the results

(e.g., Collins & Truax 1995). The price to be paid for this is the significant computational resources required, but the calculations are easily parallelized and can be performed on a reasonable time scale.

## 3. SPECTRAL ENERGY DISTRIBUTIONS

We computed a grid of NLTE plane parallel model atmospheres with PHOENIX covering temperatures between 20000 and 34000K in steps of 2000K and log g between 2.5 and 4.3 in steps of 0.3. The synthetic spectra were computed explicitly between 900 and 10000 Å. The default step size in wavelength was 0.02 Å, but PHOENIX adds wavelengths as it needs in order to adequately cover lines in NLTE which may alter the flux distribution. Some of the models have appreciable flux at wavelengths less than 900 Å, but the density of wavelengths added by PHOENIX is sufficient to capture this. Each model atmosphere provides the emergent intensity as a function of angle from the vertical at each wavelength. We also obtain the emergent flux as a function of wavelength from PHOENIX which provides a test of our integration of equation (3) for a spherically symmetric input stellar model. We find that our integration reproduces the PHOENIX flux to within 0.2 % at any given wavelength. Our integrated spectral energy distributions for these non rotating models are independent of inclination to about one part in $10^4$ and agrees with the PHOENIX flux integrated over wavelength to within about the same amount. The more interesting result is that both the integrated PHOENIX flux and our integrated flux may differ from the expected flux from the input effective temperature up to a few percent. This we attribute to the accuracy to which the temperature structure can be iterated with so many species in NLTE. One final source of error is the error associated with interpolation in the PHOENIX grid. To test this we calculated a few model atmospheres at intermediate points in the effective temperature - gravity grid and compared the PHOENIX output surface fluxes with those interpolated in the grid. We utilized linear interpolation of the log of the flux with respect to both the log of the effective temperature and of the effective gravity (the same as for the rotating models) and found this error to be about 1 to 4 % of the flux. By comparing integrated results with PHOENIX flux calculations and the known total flux for spherically symmetric stars, we found that the luminosity errors were between 1 and 6 % with the higher errors at lower luminosity.

We used the collection of PHOENIX flux spectra to devise wavelength bands which are sensitive to temperature and gravity to define color differences which could be used to determine the effective temperature. We show these flux spectra as functions of temperature for each of two gravities in Figures 1 and 2. In these figures we have convolved the flux spectra with a 20 Å wide boxcar filter and are plotting points from the resulting profile every 0.2 Å. We did not find any wavelength bands which were sufficiently sensitive to temperature alone to be used as a temperature discriminant, but we were able to find several sets of bands for which the color differences depended on gravity and temperature in sufficiently different ways that we were able to determine both adequately using two of the sets. A number of wavelength bands yielded approximately the same relationship and we chose the ones that maximized the sensitivity of the color difference to the physical variables. The first set of color differences were taken between wavelengths of 2650 - 2732 Å and 5184 - 5239 Å, and the second set covered the wavelength intervals 1972 - 2054 Å and 5506 - 5575 Å.

We show the dependence of the first set of color differences on temperature and gravity in Figure 3, and of the second set on temperature and gravity in Figure 4. In the same integral calculations we did to compare our results for a non rotating model with the PHOENIX flux, we showed that we could obtain the input effective temperature and gravity to within a degree. This should not be surprising because we used the same data to generate the relationships among the colors, gravity, and effective temperature. More interesting is the interpolation error. To test this we computed the spectral energy distributions for non rotating models which were between model atmospheres in our grid in temperature and gravity. Our temperature errors, based on the above comparison with PHOENIX flux calculations, were characteristically about 0.1 - 0.6 %, with larger errors for higher temperatures. More uncertain is the error associated with using the specific color differences for the rotating models in which the SEDs result from integrations involving a range of effective temperatures. We have performed a few tests with other wavelength intervals that show that these differences are small, but we have by no means sampled all possibilities.

## 4. INCLINATION CURVES FOR UNIFORMLY ROTATING MODELS

We have computed a suite of uniformly rotating 12 $M_\upsilon$ ZAMS models with surface equatorial velocities of 0, 50, 100, 150, 210, 255, 310, 350, 405, 450, 500, 550, and 575 km • $s^{-1}$. This last value is close to the critical rotation value of about 600 km • $s^{-1}$, so that our models cover nearly the entire range of rotation rates for uniform rotation. We present the effective temperatures as functions of colatitude for some of these models in Figure 5. As expected, the polar temperature increases slightly and the equatorial temperature decreases markedly as the rotation rate increases. The latitudinal variation of the effective temperature agrees well with that predicted by von Zeipel's (1924) law at low and intermediate rotation velocities. Both the temperature range from pole to equator and the differences between the model temperatures and those predicted by von Zeipel's law become larger for larger rotation velocities.

We show the inclination curves for these models in Figure 6. Inclinations are shown in ten degree intervals between $0^o$ and $90^o$, inclusive. Models seen pole on have the highest deduced effective temperatures and luminosities. For comparison we show the location of non rotating ZAMS models for masses of 11, 12, and 13 $M_\odot$. Clearly, the more rapidly rotating models show a great dependence between the inclination at which they would be seen and their location in the HR diagram. We note that the inclination curve as a whole has a lower average effective temperature as the rotation increases, reflecting the theoretical result that more rapidly rotating models have lower luminosities and effective temperatures.

Several other things may be noted from Figure 6. First we see that, to first order, the slope of the inclination curve is not very different from the slope of the ZAMS. This means that there need not be an obvious discriminant for rotation from the HR diagram; a rapidly rotating star may just look like a non rotating star of greater or lesser mass depending on the inclination. We also note that these results are consistent with the fact that more rapidly rotating models are cooler and less luminous (e.g., Faulkner, et al. 1968, Sackmann and Anand 1970). The total luminosity actually emitted by the model is very close to that which would be seen at an inclination very close to $55^°$. If we assume that the effects of rotation can be divided into a latitudinal dependence represented by Legendre polynomials $P_0$ and $P_2$, then the average values would be represented by the values of θ at which $P_2$ goes to zero (approximately 55°).

We compare our inclination curves with those generated using von Zeipel's (1924) law in Figure 7. Von Zeipel's law only provides a proportionality between the effective temperature and the effective gravity, so we must provide the proportionality constant. We could do this at any latitude because we compute both the effective gravity and effective temperature in each angular zone at the surface. We have chosen to do this at the angular zone closest to θ = 55°. This colatitude was chosen for the reason given in the previous paragraph. The actual colatitude used is 58.5° which is the midpoint of one of our angular zones. While not exact, the intent is for the two models to have approximately the same luminosity. At low rotation velocity our results and von Zeipel's law agree quite closely, but begin to differ as the difference between the polar and equatorial effective temperatures becomes significant. LDS explained this as an implicit decoupling of the surface temperature and the effective temperature by von Zeipel's law. With von Zeipel's law the surface temperature must be constant because the surface is an equipotential, while the effective temperature varies significantly from pole to equator. When the surface equatorial velocity is 350 km • $s^{-1}$ the effect is noticeable but von Zeipel's law still provides a reasonable representation. This is progressively less so at high rotational velocities, with the von Zeipel effective temperatures being higher at the pole and lower at the equator than the temperatures from our models. The luminosity at the pole remains about the same for the two calculations, but the von Zeipel luminosity becomes significantly lower at latitudes less than about 30°.

The general shape of our inclination curves agree with those of other authors (e.g.,Hardorp & Strittmatter 1968; Collins & Sonneborn 1977; Collins, Truax & Cranmer 1991) for models of this approximate mass: the inclination curves move to the right on the HR diagram as the rotation rate increases but remain approximately parallel to the ZAMS. They also get longer as the rotation rate increases. Approaches which assume that the polar temperature and radius are constant with increasing rotation do not move as far to the right or down in the HR diagram as do models which include the effects of rotation on the stellar model (e.g., Collins, Truax & Cranmer 1991). Detailed comparisons of inclination curves with previous calculations are much more difficult because most of the information is given only as a curve on an HR diagram.

## 5. INCLINATION CURVES FOR DIFFERENTIALLY ROTATING MODELS

We have computed several differentially rotating ZAMS models. The rotation law is conservative, given by the equation

$$\Omega(\varpi) = \frac{\Omega_0}{1 + (a\varpi)^\beta} \tag{5}$$

In this equation, $\Omega$ is the rotation rate, $\varpi$ is the distance from the rotation axis (scaled to be unity at the equator), and $\Omega_0$, a, and β are constants used to impose the surface rotational velocity, the distance from the rotation axis at which the differential rotation becomes significant, and the amount of differential rotation. The rotation laws which can most affect the model structure are those in which the rotation rate in the deep interior is high, i.e., β>0. Stability considerations require β<2, so we will consider here only cases with values of β between these two values. We have computed two sets of models, both at 10 $M_\odot$. The first set has a surface equatorial velocity of 120 km • $s^{-1}$ with values of β between 0 and 2 in steps of 0.2. The second set covers the same values of β, but with a surface equatorial velocity of 240 km • $s^{-1}$. We have arbitrarily chosen a = 2. We show the effective temperature as a function of colatitude from the ROTORC models with the higher surface equatorial velocity for selected values of β in Figure 8. Note that the net effect of increasing β is to increase the range between the polar and equatorial effective temperatures in much the same way that increasing the rotation rate for uniform rotation does. This same trend is noticeable in the change in the surface shape, seen in Figure 9, where increasing β decreases the ratio of the surface polar radius to the surface equatorial radius. In both cases the detailed profiles differ between increasing β and increasing the uniform rotation rate, but the general trend is the same. We do not consider higher values of β because the surface radius increases so sharply near the pole for high β that someone standing near the pole would actually see part of the surface above the horizon. This could be accommodated at the pole by the plane parallel approximation with incident radiation from outside the atmosphere, But not near the pole where the incident radiation above the horizon would not be azimuthally symmetric about the local normal to the surface. We should not use our plane parallel atmosphere approach to determine the spectral energy distributions for this extreme model.

We have computed spectral energy distributions and inclination curves for two differentially rotating models. These have the surface equatorial velocity of 240 km • s$^{-1}$ and values of β of 0, 0.6, and 1.2. The latter is the highest value for which the effective temperatures at all latitudes all fit within our model atmosphere grid. The inclination curves for the three models are presented in Figure 10. Given the greater range between the polar and equatorial temperatures, it should not be a surprise that the inclination curves are longer for the differentially rotating models. While the slope of the inclination curves has changed for the differentially rotating models, it is still approximately parallel to the ZAMS. One expects that if one knows only the observed location of a particular massive star in the HR diagram, one would be hard pressed to deduce anything about its internal or even global properties.

## 6. CONCLUSIONS

We have performed stellar structure and model atmosphere calculations to determine the spectral energy distributions for both uniformly and differentially rotating stars. These spectral energy distributions were used to determine what an observer would deduce the effective temperature and luminosity to be assuming the star was not rotating. These deductions would depend on the inclination between the polar axis and the direction to the observer. The variation of these deduced quantities from pole to equator can be large, and one would make serious errors in the estimated values obtained assuming spherical stars once the rotational velocities are over approximately 300 km • s$^{-1}$ for our 12 $M_\odot$ models. This velocity corresponds to a rotation rate of about 0.7 of critical rotation. Any attempt to place rapidly rotating stars in an HR diagram must account for the variation of the fundamental properties of brightness and color as a function of inclination.

The inclination curves for nonuniformly rotating stars differ somewhat from those of uniformly rotating stars, particularly as the rotation rate increases. This adds to the complexity of deducing reliable luminosities and effective temperatures for rotating stars, but has the potential, when combined with other information, to allow one to deduce information about the angular momentum distribution inside rapidly rotating stars.

This work is supported by Discovery Grants to Drs. Deupree and Short from the National Sciences and Engineering Research Council of Canada (NSERC) and by a NSERC graduate fellowship to Ms. Lovekin. Computational facilities are provided in part by ACEnet, the regional high performance computing consortium for universities in Atlantic Canada. ACEnet is funded by the Canada Foundation for Innovation (CFI), the Atlantic Canada Opportunities Agency (ACOA), and the provinces of Newfoundland & Labrador, Nova Scotia, and New Brunswick. The authors are grateful for this support.

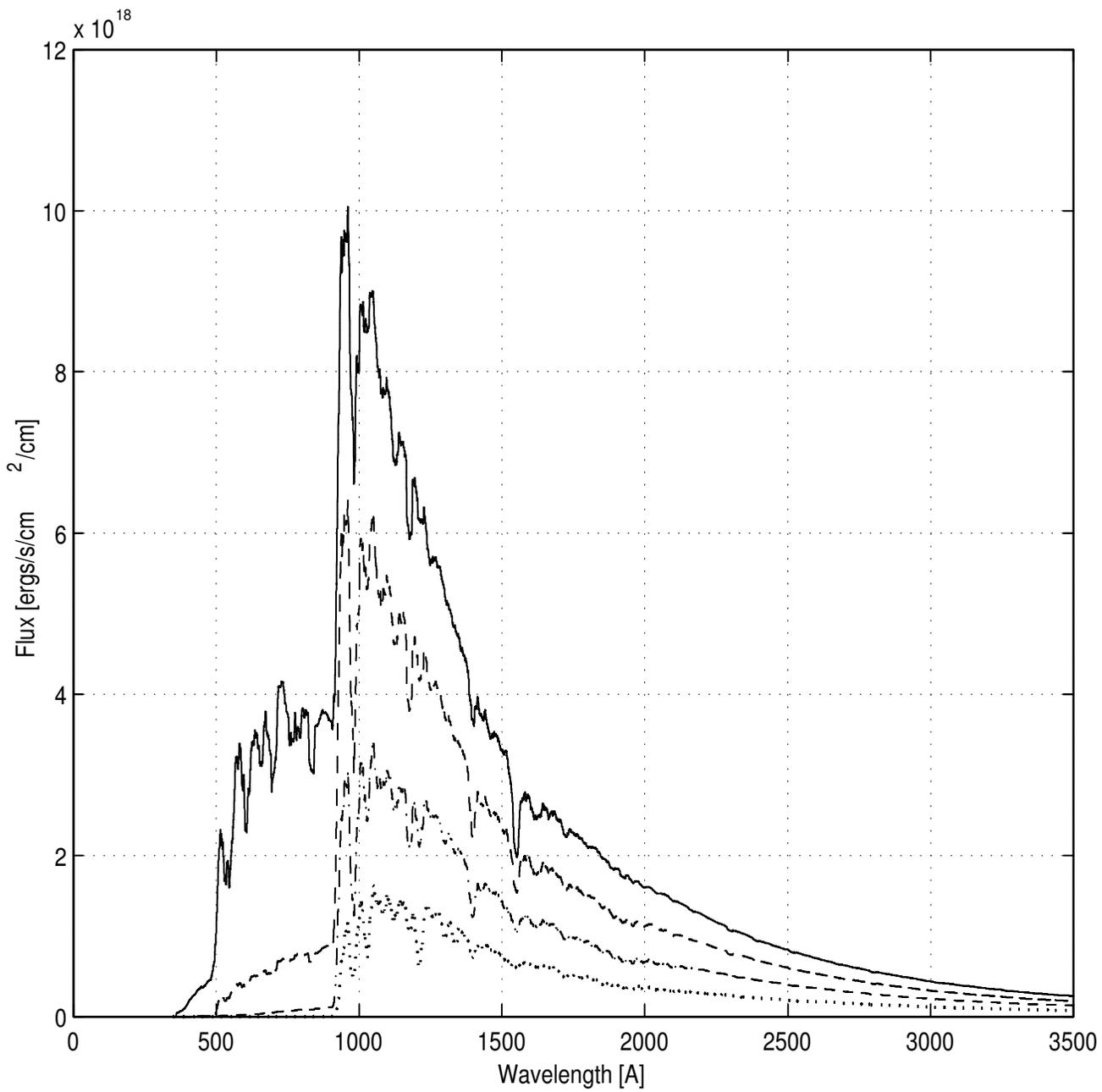

Fig. 1. - Emergent fluxes from PHOENIX plane parallel model atmospheres for log(g) = 3.4 for effective temperatures from 22000K (bottom) to 34000K (top) in increments of 4000K. The actual grid has effective temperatures twice as dense as this.

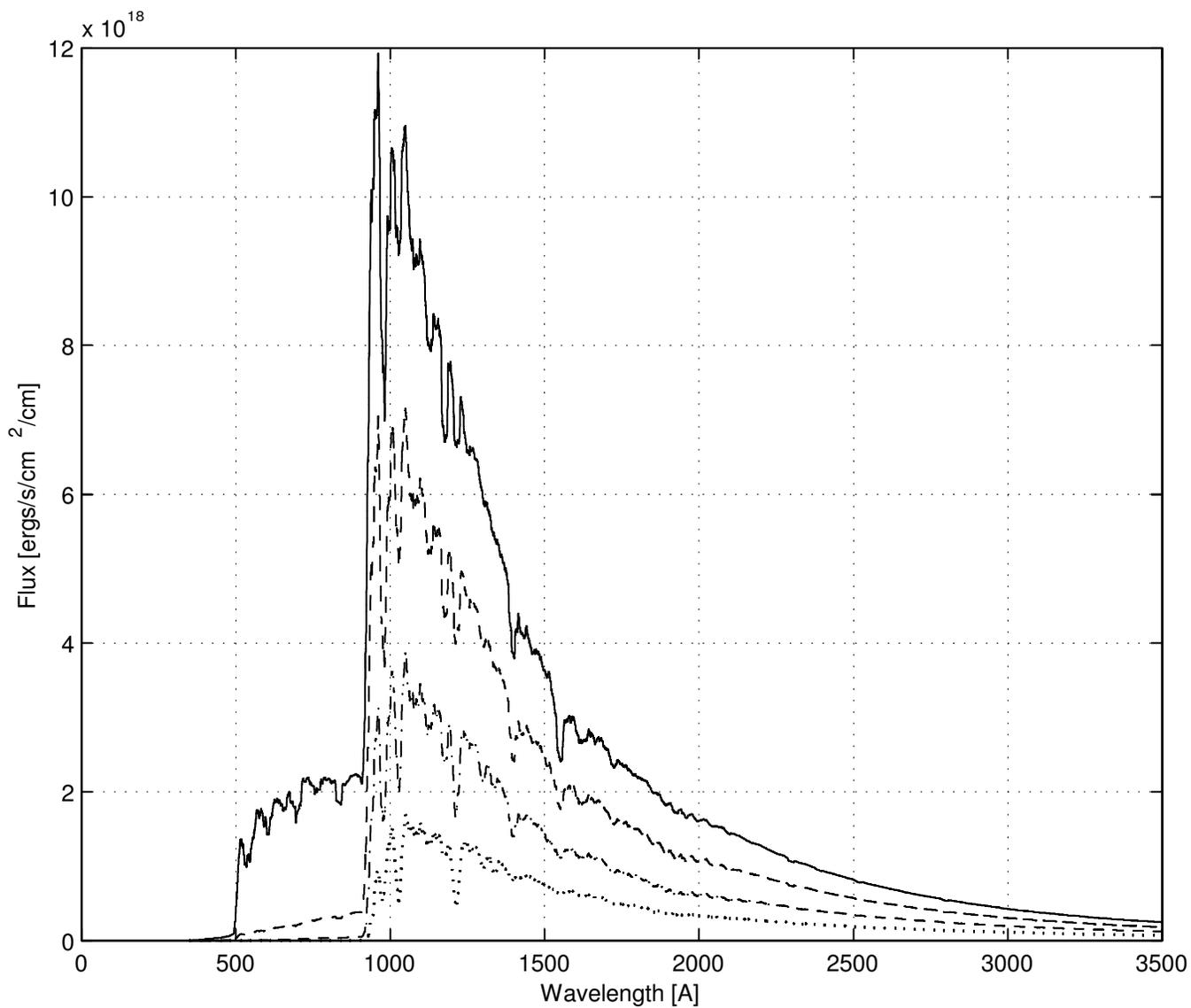

Fig. 2. - Emergent fluxes from PHOENIX plane parallel model atmospheres for log(g) = 4.3 for effective temperatures from 22000K (bottom) to 34000K (top) in increments of 4000K. The actual grid has effective temperatures twice as dense as this.

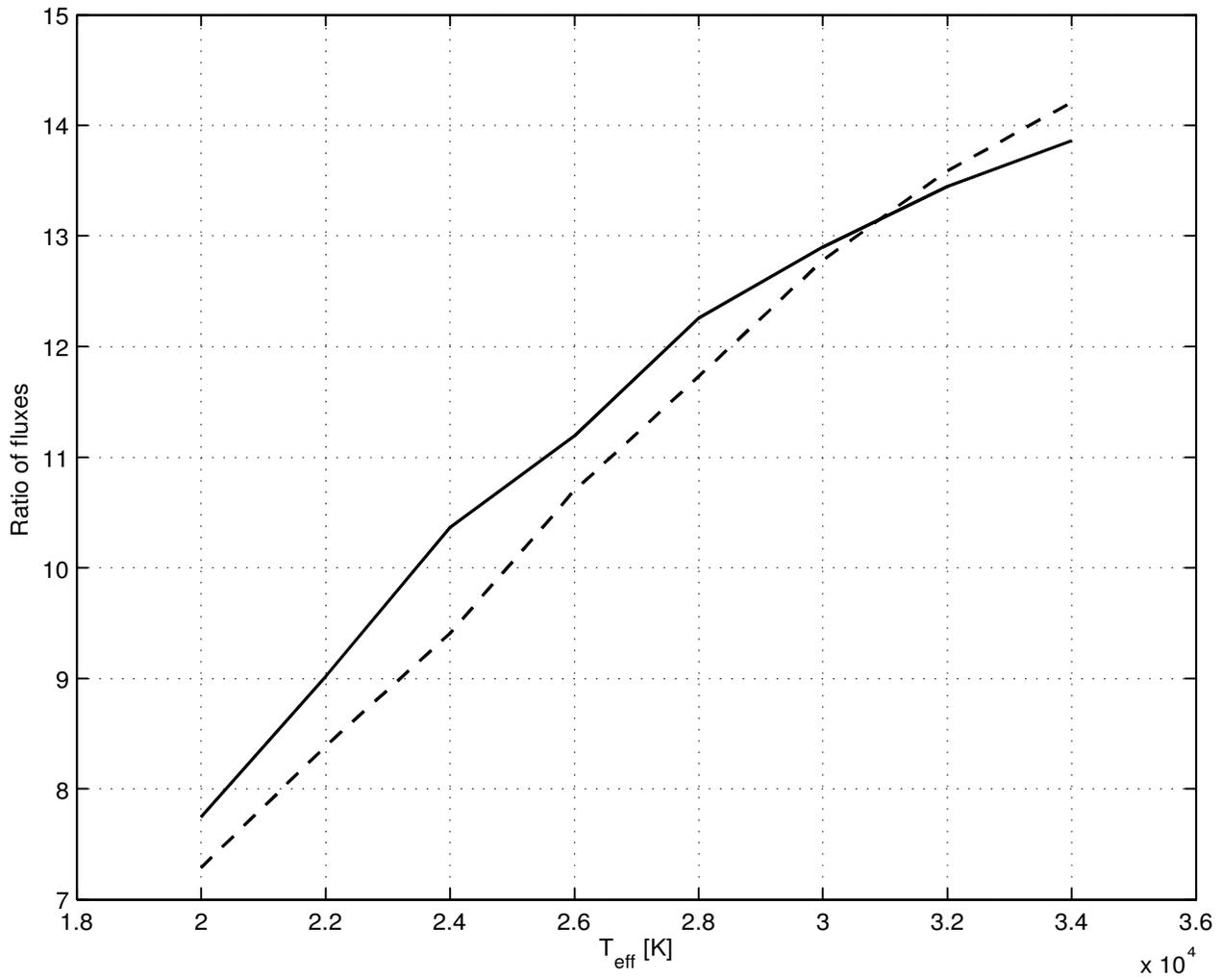

Fig. 3. - Ratio of the flux in the wavelength interval 2650 - 2732 Å to the flux in the wavelength interval 5184 - 5239 Å as a function of temperature. This ratio is computed from the integrated spectral energy distributions for nonrotating models. The solid curve is for a log(g) of 4.3 and the dashed curve for log(g) of 3.4.

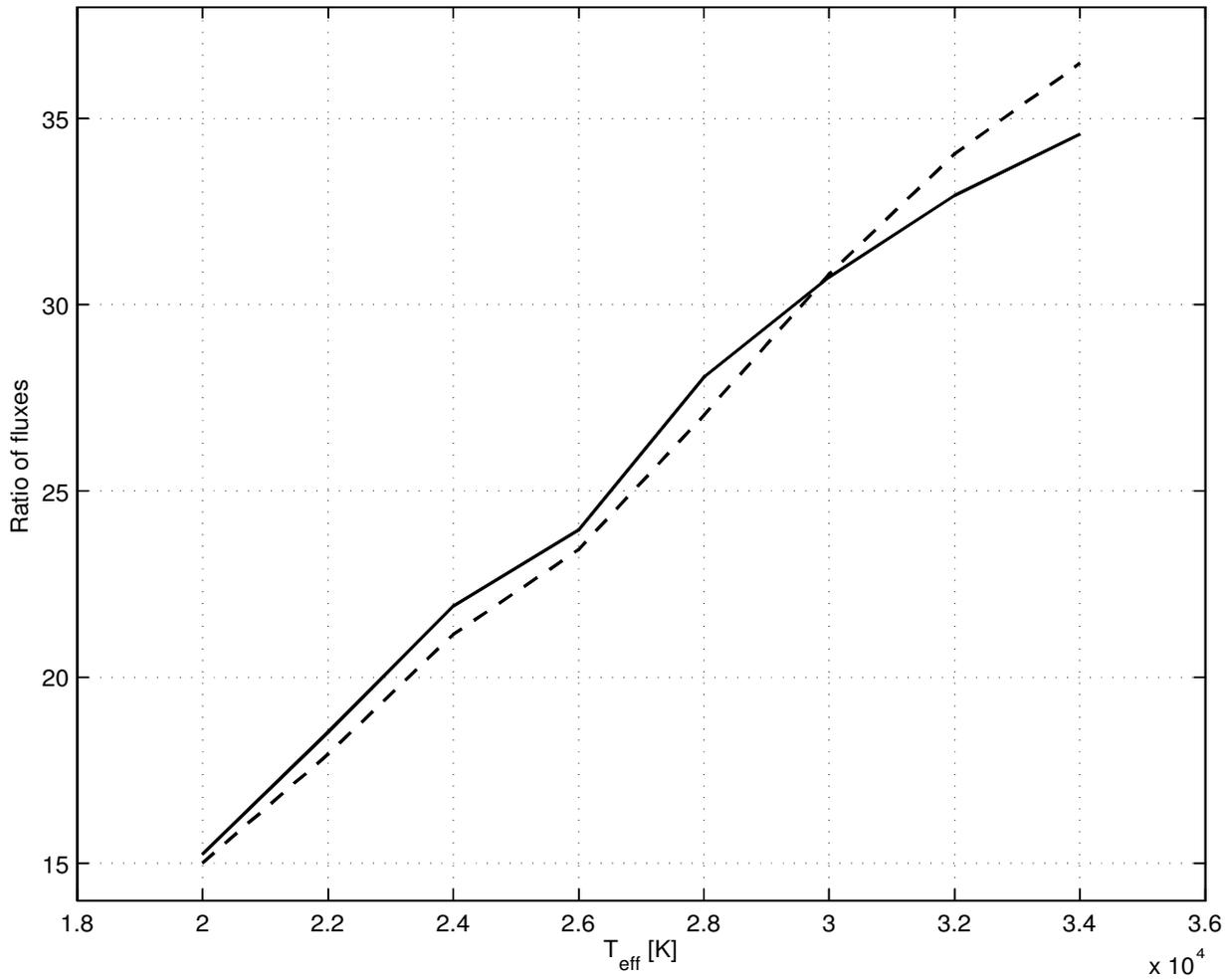

Fig. 4. - Ratio of the flux in the wavelength interval 1972 - 2054 Å to the flux in the wavelength interval 5506 - 5575 Å as a function of temperature. This ratio is computed from the integrated spectral energy distributions for nonrotating models. The solid curve is for a log(g) of 4.3 and the dashed curve for log(g) of 3.4.

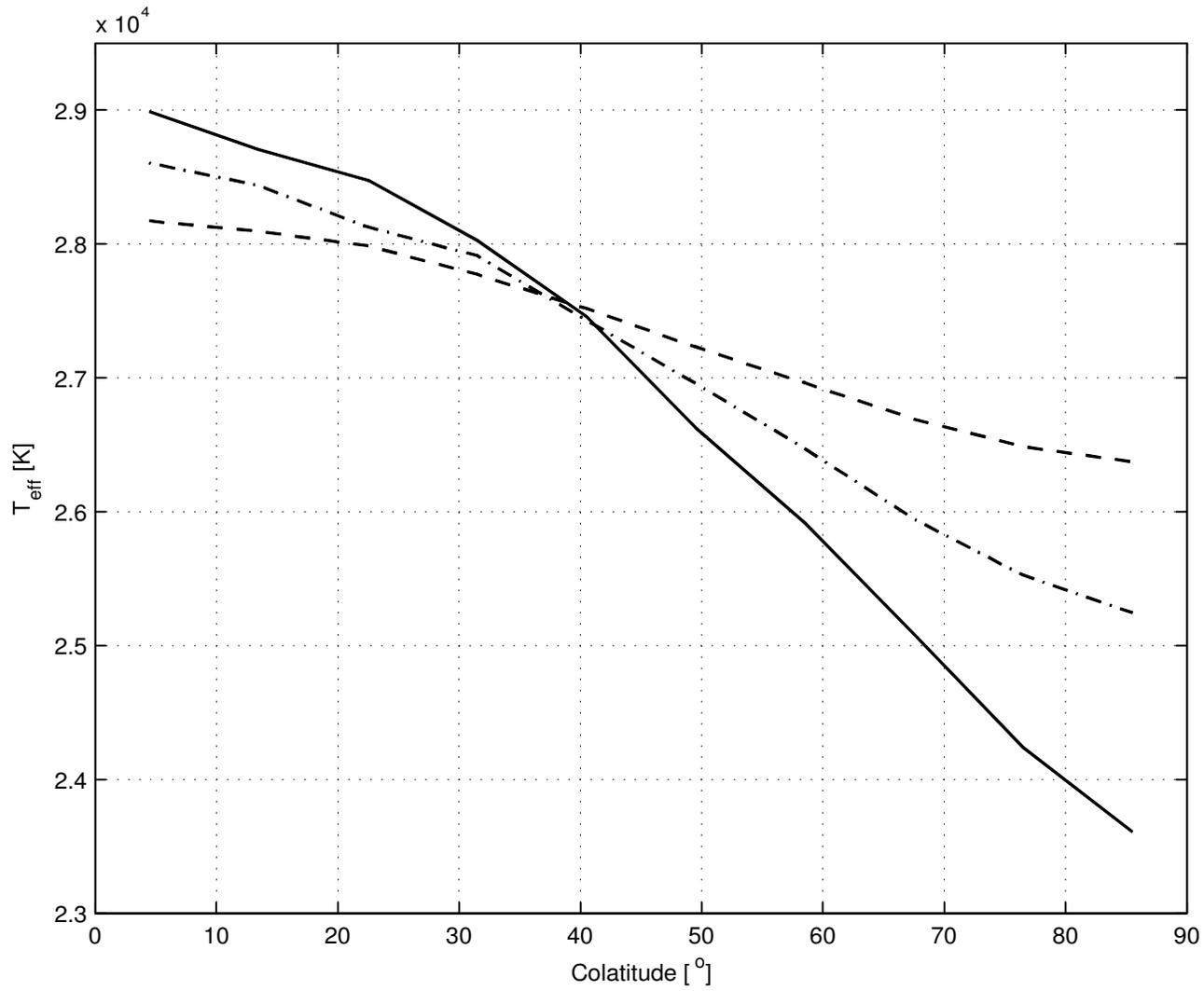

Fig. 5. - Effective temperature versus colatitude for selected uniformly rotating models taken from the 2D stellar structure calculations. All are 12 $M_\odot$ ZAMS models with Z=0.02. The polar temperature increases and the equatorial temperature decreases with increasing rotation rate. The surface equatorial velocity for the models are: 255 (dashed curve), 350 (dot dashed curve), and 450 (solid curve) km • $s^{-1}$.

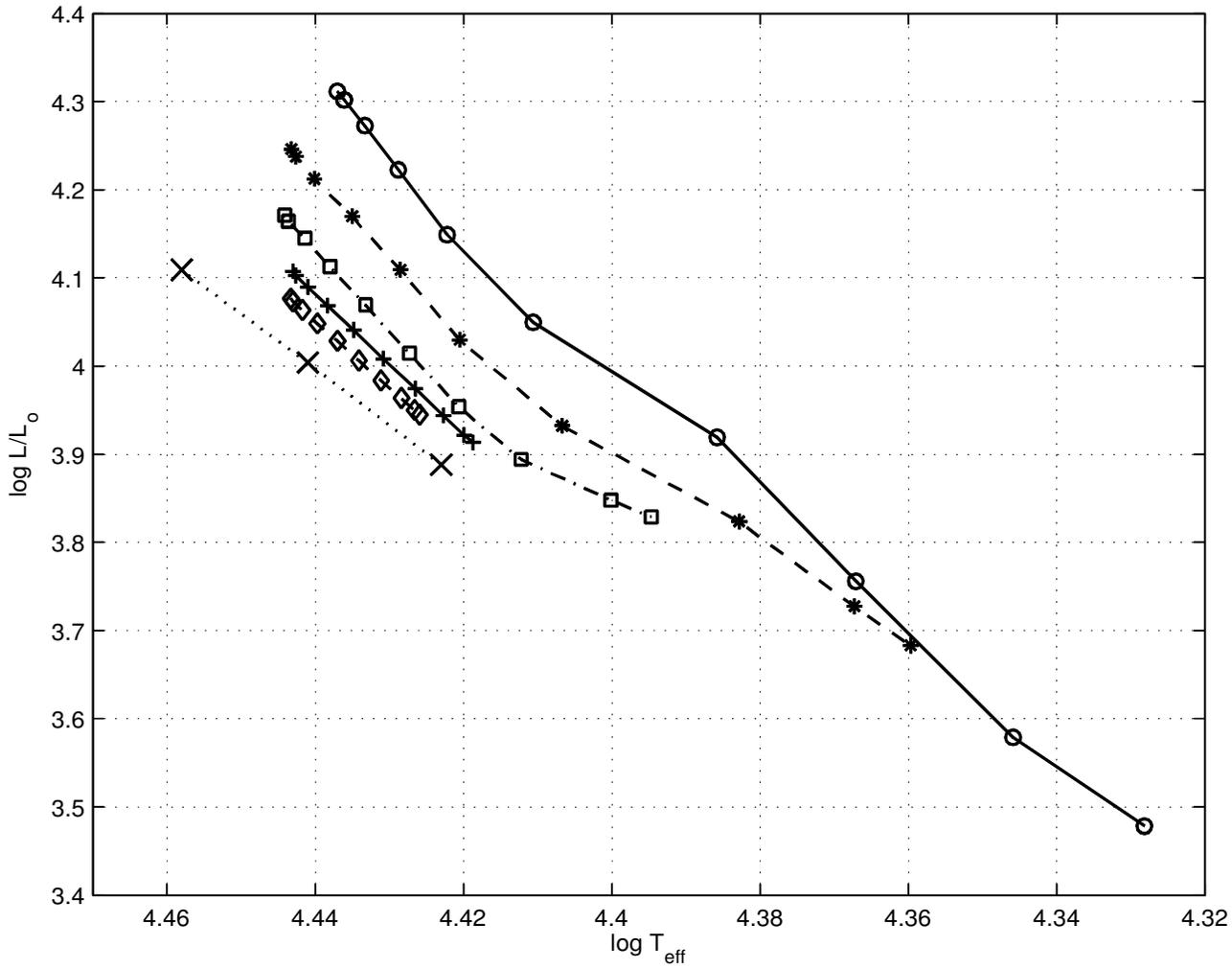

Fig. 6. - Inclination curves for a selection of uniformly rotating ZAMS models. The deduced luminosity and effective temperature for a given inclination is determined by the area under the SED and by the shape of the SED, respectively. The inclination curves are longer for more rapidly rotating models. The surface equatorial velocities for these models are 255 (◊), 310 (+), 405 (□), 500 (*), and 575 (o) km • s$^{-1}$. The large X's are the locations of nonrotating ZAMS models with masses of 11, 12, and 13 M$_\odot$.

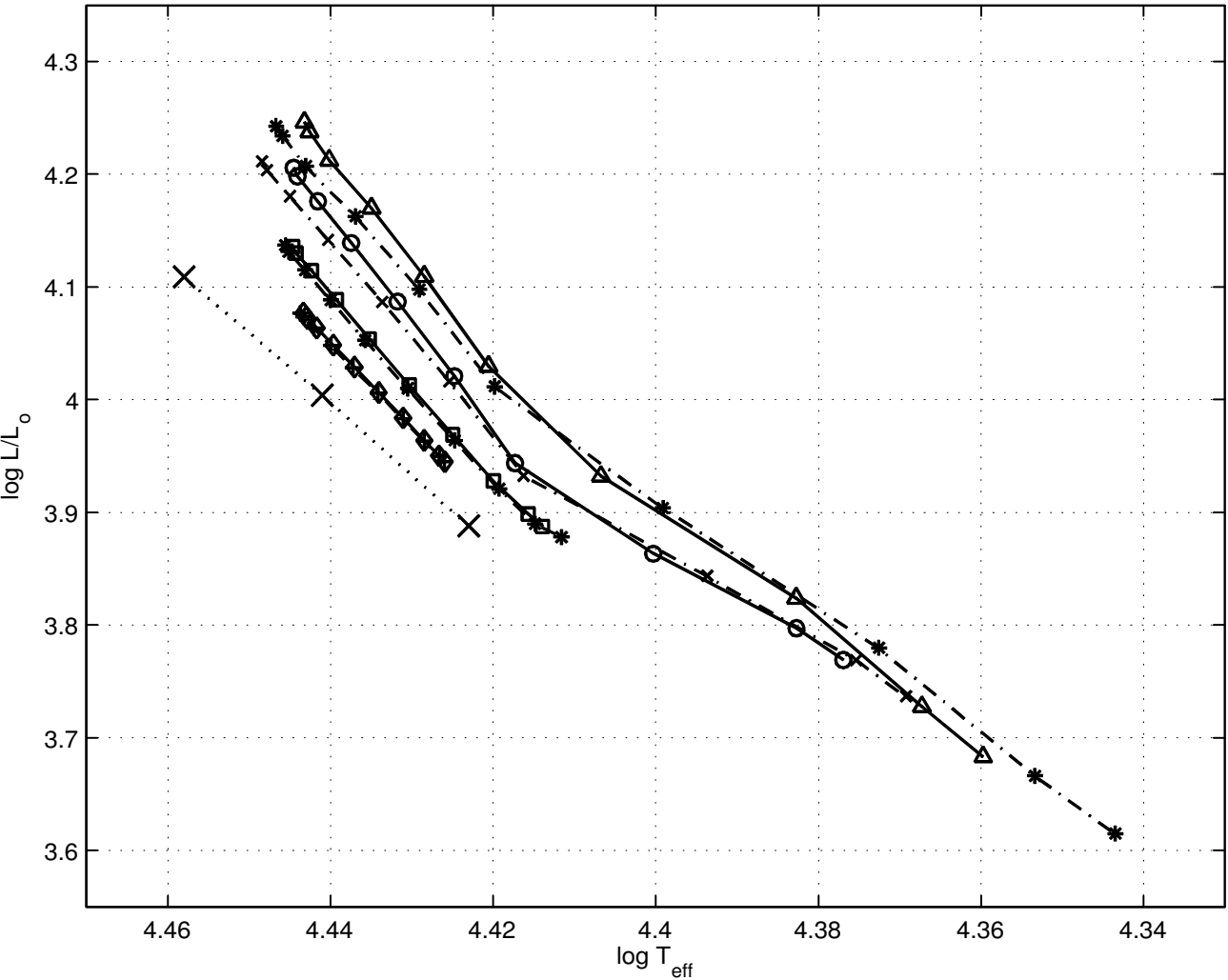

Fig. 7. - Comparison of the inclination curves obtained from a full 2D ZAMS stellar model (solid curves) with those obtained from von Zeipel's law (dot dashed curves) for selected rotation rates. The curves are presented for surface equatorial velocities of 255 (◊, +), 350 (□, *), 450 (o, x), and 500 (△, *) km • s$^{-1}$. The first symbol refers to the ROTORC model and the second to the von Zeipel model. The large X's are the locations of nonrotating ZAMS models with masses of 11, 12, and 13 M$_\odot$. The two sets of inclination curves agree at small and moderate rotation velocities, but are significantly different at higher rotation velocities.

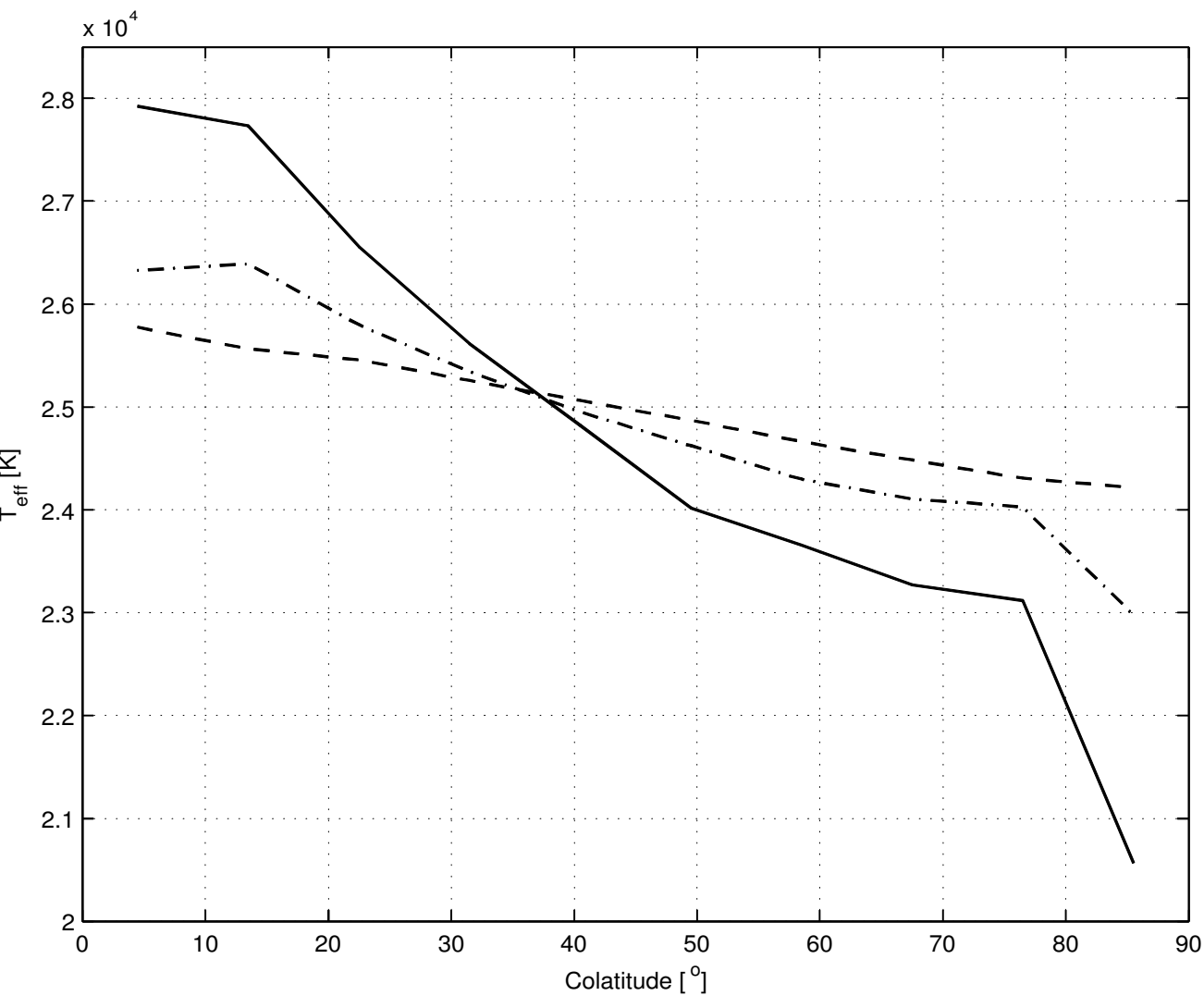

Fig. 8. - Effective temperature as a function of colatitude for differentially rotating models with a surface equatorial velocity of 240 km • s$^{-1}$. The differential rotation is specified by the parameter β as given in equation (5). The values of β in this figure are 0 (dashed curve), 0.6 (dot dashed curve), and 1.2 (solid curve).

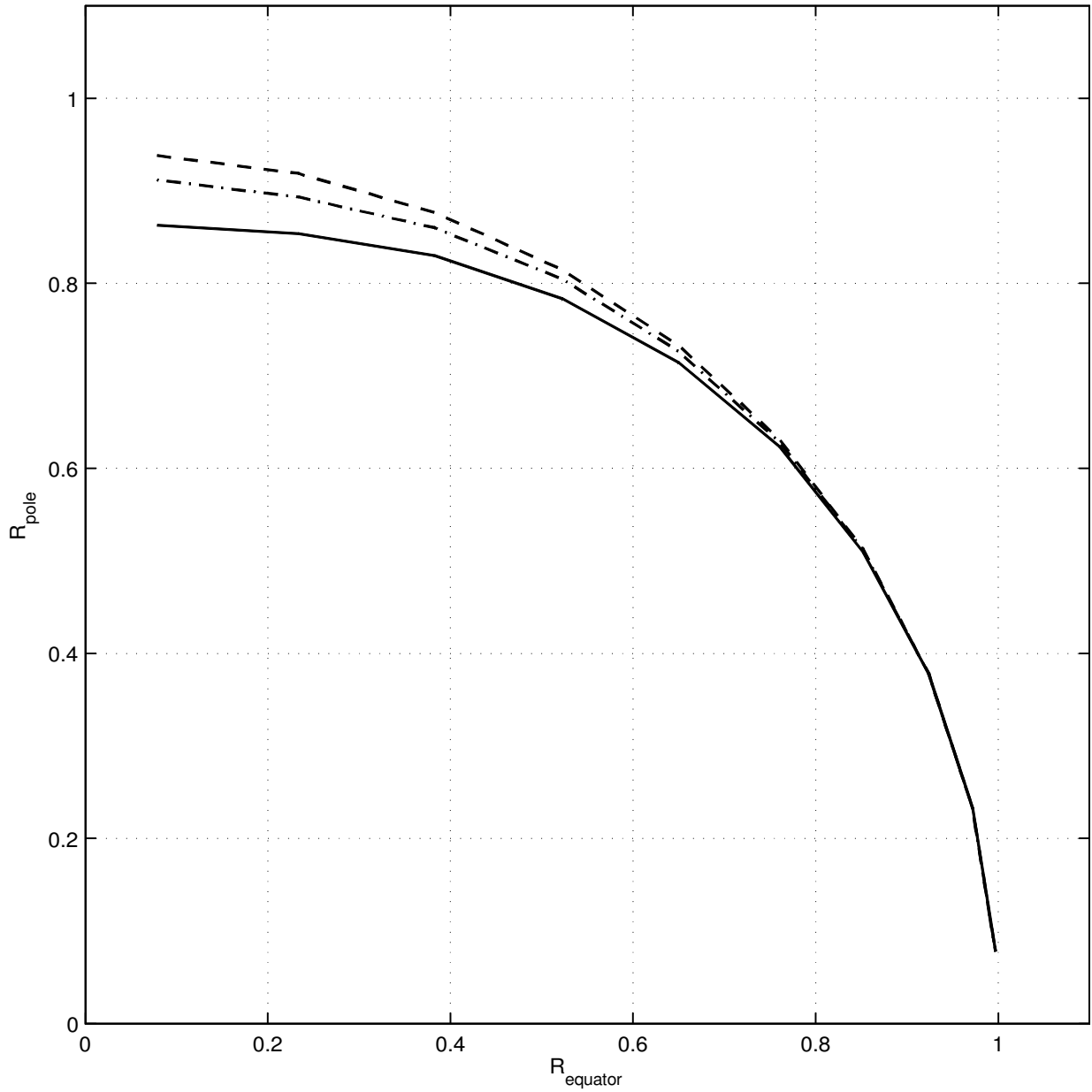

Fig. 9. - The surface location, scaled to the surface equatorial radius, for three differentially rotating models, all with a surface equatorial velocity of 240 km • s$^{-1}$. The differential rotation is specified by the parameter β as given in equation (5). The values of β in this figure are 0 (dashed curve), 0.6 (dot dashed curve), and 1.2 (solid curve). Higher values of β produce smaller polar radii.

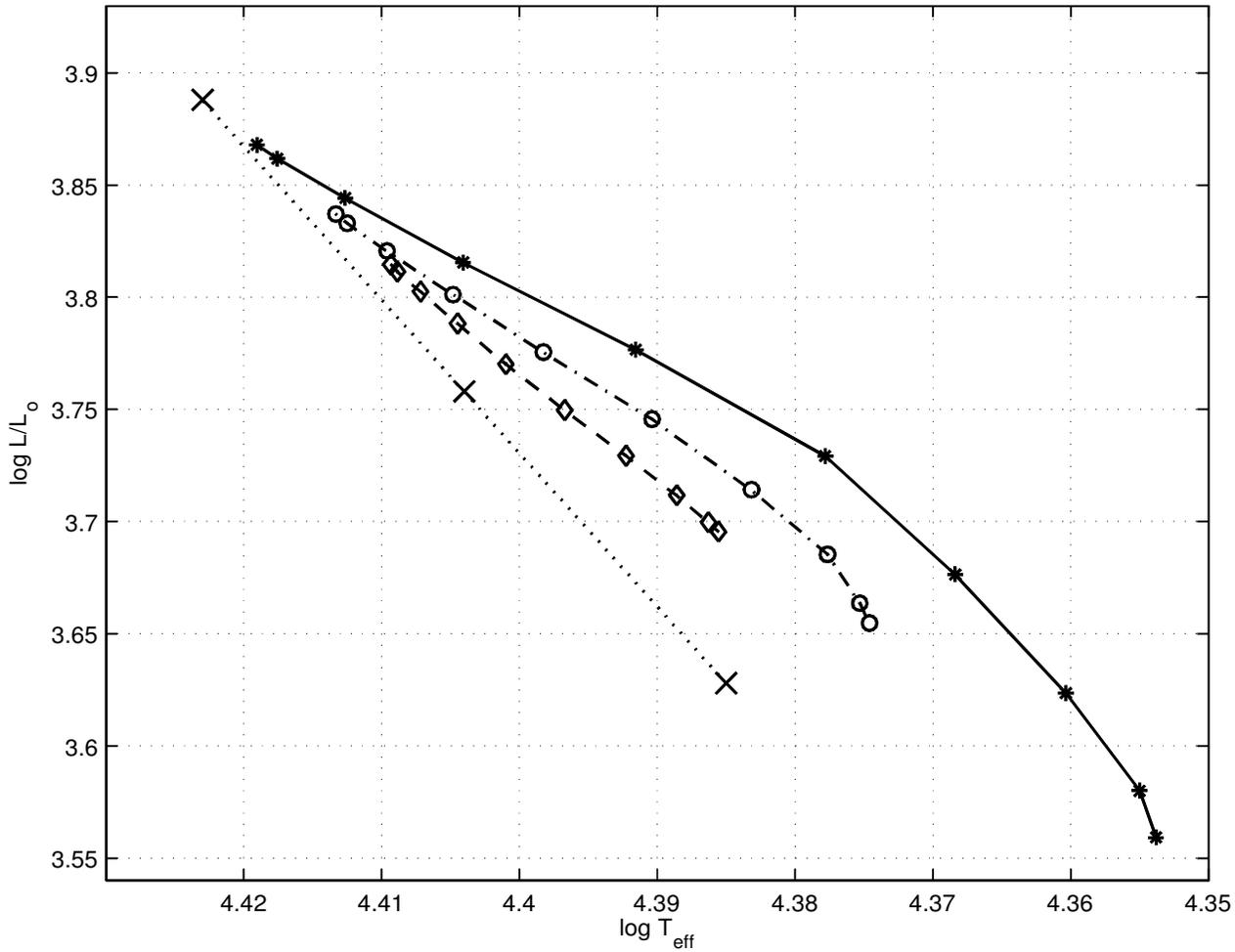

Fig. 10. - Inclination curves for three differentially rotating models, all with a surface equatorial velocity of 240 km •s$^{-1}$. The differential rotation is specified by the parameter β as given in equation (5). The values of β in this figure are 0 (◊), 0.6 (o), and 1.2 (*). The large X's denote the location of nonrotating ZAMS models at the masses given. Greater differential rotation both lengthens and slightly changes the slope of the inclination curve.

TABLE 1
SPECIES/LINES TREATED IN NLTE

| Element | Ionization Stage | | | |
|---|---|---|---|---|
| | I | II | III | IV |
| H......... | 50/1255 | ........ | ....... | ......... |
| He........ | 19/37 | 10/45 | ......... | ......... |
| Li......... | 57/333 | 55/124 | ..... | ..... |
| C.......... | 228/1387 | 85/336 | 79/365 | ..... |
| N.......... | 252/2313 | 152/1110 | 87/266 | ..... |
| O.......... | 36/66 | 171/1304 | 137/765 | ..... |
| Ne........ | 26/37 | ..... | ..... | ..... |
| Na........ | 53/142 | 35/171 | ..... | ..... |
| Mg........ | 273/835 | 72/340 | 91/656 | ..... |
| Al......... | 111/250 | 188/1674 | 58/297 | 31/142 |
| Si......... | 329/1871 | 93/436 | 155/1027 | 52/292 |
| P.......... | 229/903 | 89/760 | 51/145 | 50/174 |
| S.......... | 146/439 | 84/444 | 41/170 | 28/50 |
| K.......... | 73/210 | 22/66 | 38/178 | ..... |
| Ca......... | 194/1029 | 87/455 | 150/1661 | ..... |
| Fe......... | 494/6903 | 620/13675 | 566/9721 | 243/2592 |